\documentclass[prd,twocolumn,superscriptaddress,floatfix,amsmath,amssymb,amsfonts,nofootinbib,longbibliography]{revtex4-2}

\usepackage{float} 
\usepackage{needspace}
\usepackage{scalerel}
\usepackage[normalem]{ulem}
\usepackage[english]{babel}
\usepackage{graphicx}
\usepackage{dcolumn}
\usepackage{bm}
\usepackage{blindtext}
\usepackage{verbatim}
\usepackage{relsize}
\usepackage{mathrsfs}
\usepackage{musicography}
\usepackage{amsmath}
\usepackage{blindtext}
\usepackage{cancel}
\usepackage{physics}
\usepackage{epstopdf}
\usepackage{mathtools}
\usepackage{blindtext}
\usepackage{tensor}
\usepackage{color}
\usepackage[usenames,dvipsnames]{pstricks}
\usepackage{epsfig}
\usepackage{pst-grad} 
\usepackage{pst-plot} 
\usepackage{hyperref}
\usepackage{verbatim}
\usepackage{slashed}
\usepackage{dsfont}

\hypersetup{
    colorlinks=true,
    linkcolor=blue,
    filecolor=red,      
    urlcolor=cyan,
}

\newcommand{\ii}{\mathrm{i}}

\usepackage[normalem]{ulem}

\allowdisplaybreaks[1] 

\begin{document}

\title{Zero mode, Zero influence: Spatial-derivative coupling in particle detectors}

\author{Eduardo Mart\'{i}n-Mart\'{i}nez}
\email{emartinmartinez@uwaterloo.ca}
\affiliation{Department of Applied Mathematics, University of Waterloo, Waterloo, Ontario, N2L 3G1, Canada}
\affiliation{Institute for Quantum Computing, University of Waterloo, Waterloo, Ontario, N2L 3G1, Canada}
\affiliation{Perimeter Institute for Theoretical Physics, Waterloo, Ontario, N2L 2Y5, Canada}
\affiliation{Department of Physics and Astronomy, University of Waterloo, Waterloo, Ontario, N2L 3G1, Canada}

\author{Ireneo James Membrere}
\email{imembrere@uwaterloo.ca}
\affiliation{Department of Applied Mathematics, University of Waterloo, Waterloo, Ontario, N2L 3G1, Canada}
\affiliation{Institute for Quantum Computing, University of Waterloo, Waterloo, Ontario, N2L 3G1, Canada}
\affiliation{Perimeter Institute for Theoretical Physics, Waterloo, Ontario, N2L 2Y5, Canada}

\author{Adam Teixid\'{o}-Bonfill}
\email{adam.teixido-bonfill@uwaterloo.ca}
\affiliation{Institute for Quantum Computing, University of Waterloo, Waterloo, Ontario, N2L 3G1, Canada}
\affiliation{Department of Physics and Astronomy, University of Waterloo, Waterloo, Ontario, N2L 3G1, Canada}

\begin{abstract}
It is common in particle detector models in cavity QED  to often omit zero modes or assume their state, compromising causality or introducing ambiguity in their predictions. We show that spatial-derivative coupling eliminates both problems. For free massless scalar fields in static periodic or Neumann cavities of any spatial dimension, cavity-frame spatial derivatives annihilate the zero-mode operator while preserving microcausality. Detector dynamics and entanglement harvesting then depend only on the reduced detector–oscillator state, nonperturbatively and even with initial correlations with the zero mode. Time derivatives and moving-detector rest-frame spatial derivatives can retain zero-mode state dependence not compromising causality. Realized through current coupling in superconducting circuits, spatial-derivative coupling thus makes zero-mode state assumptions unnecessary.
\end{abstract}

\maketitle


\Needspace{6\baselineskip}
\section{Introduction}

Particle detector models provide an operational description of how localized quantum systems probe a quantum field. The Unruh--DeWitt (UDW) model describes a two-level system coupled linearly to a scalar field \cite{Unruh1976_blackholeevap,DeWitt1980_UDWDetector,UnruhWald1984_AcceleratingObserverRindlerParticle}, and has been used to study entanglement harvesting \cite{Valentini1991_NonlocalCorrelationsQED,Reznik2003_EntanglementVacuum,Pozas2015_HarvestingCorrelationsQuantumVacuum}, communication through quantum fields \cite{Cliche2010_RelativisticQuantumChannel,Jonsson2014_QuantumSignallingCavityQED,Jonsson2015_InformationTransmission}, and the Unruh effect~\cite{Crispino2008_UnruhEffectReview}. In cavity implementations and models with periodic spatial sections, the boundary conditions determine which field modes contribute to these processes. A description that retains only oscillator modes therefore requires examining the physical consequences of any omitted degrees of freedom.

For a free massless scalar field in a static cavity with periodic or Neumann boundary conditions in all spatial directions, this issue arises from the spatially homogeneous zero mode. Its dynamics are those of a quantum free particle, and its Hamiltonian does not select a normalizable ground state that could play the role of a Fock vacuum\footnote{Zero-mode obstructions to vacuum construction also arise in curved spacetime. For a massless minimally coupled scalar field in de Sitter spacetime, the constant zero mode in the Euclidean construction makes the massless limit of the field two-point function divergent. Page and Wu obtain well-defined de Sitter-invariant correlations by restricting to observables invariant under constant field shifts, for which this divergent contribution cancels \cite{Page2012_ShiftInvariantVacuum}.}\cite{MartinMartinez2014_ParticleDetZeroMode,Yazdi2017_ZeroModesEntropy}. This obstruction has motivated calculations in which the zero mode is removed by hand. As discussed in Refs.~\cite{Tjoa2019_ZeroModeSuperluminalSignal,Tjoa2020_ETZeroMode}, such omissions can be found in studies of entanglement harvesting \cite{Brown2013_thermalAmplificationEH,Lorek2014_TripartiteEH}, detector thermometry \cite{Robles2017_QuantumThermometry}, and accelerated detectors \cite{BRENNA2016_AntiUnruh}. The absence of a Fock-vacuum prescription, however, does not imply that the zero mode has a negligible effect on these observables.

The consequences of this omission go beyond changing numerical predictions. Removing the zero mode produces an acausal retarded Green's function on the Einstein cylinder~\cite{Lin2016_EntanglementEinsteinCylinder}. For detectors coupled to the field amplitude, this permits signalling between spacelike-separated interaction regions~\cite{Tjoa2019_ZeroModeSuperluminalSignal}. In a study of spacelike entanglement harvesting, this introduces a communication channel that should be absent and compromises the interpretation of the extracted correlations. The zero mode is needed to recover the microcausality of the amplitude field. Its exclusion therefore cannot be justified solely as a convenient way of applying the oscillator formalism.


Including the zero mode does not, by itself, supply a unique prediction for detector dynamics. The choice of the oscillator vacuum leaves the zero-mode state unspecified. Previous work showed that this state affects detector responses \cite{MartinMartinez2014_ParticleDetZeroMode} and can substantially change the entanglement harvested by two detectors at leading order in their couplings \cite{Tjoa2020_ETZeroMode}. Consequently, calculations described as harvesting from the vacuum can give different results for the same oscillator vacuum and detector parameters, depending on the chosen zero-mode state. This is the ambiguity identified in Ref.~\cite{Tjoa2020_ETZeroMode}: without a physical prescription for that state, the vacuum condition on the oscillator sector does not determine the harvesting prediction. The dependence persists for time-derivative coupling, although its contribution can be made small for suitable states and in appropriate limits \cite{MartinMartinez2014_ParticleDetZeroMode,Tjoa2020_ETZeroMode}. Such suppression still requires a reason to select the state or regime in question. For example, a geometrically motivated prescription for the zero-mode state was developed in~\cite{AlonsoSerrano2021_ZeroModeQuantization}  relating the quantization of the zero mode in the Einstein cylinder to a family of time-machine geometries.


These results raise the question of whether a physically motivated detector interaction can avoid dependence on the zero-mode state altogether. Spatial-derivative coupling is a promising candidate: since the zero mode is spatially homogeneous, the spatial derivative may remove it from the interaction. Such couplings also occur naturally in superconducting circuits. A flux qubit inductively coupled to a transmission line probes the line's current, which is proportional to the spatial derivative of the line's flux, a scalar field. In a discrete circuit description, the corresponding interaction involves a difference between neighbouring node fluxes \cite{peropadre2013PRLSuppl}. Beyond this theoretical description, tunable coupling between a flux qubit and the current in a transmission line has been demonstrated experimentally \cite{Janzen2023_TunableCouplerTLSWaveguide}. Therefore, spatial-derivative coupling is a physically realistic interaction with which to examine independence from the zero-mode state.

Here we show that detectors coupled only to spatial derivatives of a free massless scalar field are exactly independent of the zero-mode state in static rectangular cavities with periodic or Neumann boundary conditions in any spatial dimension. We first give an explicit calculation on the static $(1+1)$-dimensional cylinder and then establish the general result. Every spatial derivative annihilates the homogeneous zero-mode operator, so the oscillator sector gives the same detector dynamics as the full field, with the initial reduced state of the detectors and oscillator sector held fixed. This is an identity at the level of the interaction and does not require a special zero-mode state, a limiting procedure, or a perturbative truncation. We also establish that the coupled field observable remains microcausal. Comparison with time-derivative coupling separates these two properties: both derivative--derivative commutators vanish at spacelike separation, but only the spatial derivative eliminates the zero-mode contribution to the interaction. Within these models, spatial-derivative coupling therefore removes the additional state specification responsible for the ambiguity in earlier detector and harvesting calculations.


In Sec.~\ref{sec:review}, we review the zero-mode dynamics and its effects on detector responses, signalling, and entanglement harvesting. Section~\ref{sec:results} establishes the causal behaviour of the derivative observables and the exact decoupling of the spatial-derivative interaction from the zero mode on the static cylinder. Section~\ref{sec:arbitrary_dimensions} extends these results to any spatial dimension for static periodic and Neumann cavities. We discuss the implications in Sec.~\ref{sec:conclusions}. We use natural units, $\hbar=c=1$.

\Needspace{6\baselineskip}
\section{Review of the Zero Mode}
\label{sec:review}

Recall that zero modes occur for massless scalar fields defined on a Minkowski spacetime confined to a static cavity with periodic boundary conditions. In this section, we briefly outline the mathematical formalism for defining a zero mode in this case. Then we summarize previous works to highlight two key issues that arise for particle detector models coupled to these fields: the ambiguity in choosing the zero-mode state, and violations of causality.

\subsection{Scalar Quantum Field Confined to a Cavity}
Consider a massless scalar quantum field defined on a $1+1$-dimensional Minkowski spacetime that is confined to a cavity of length $L$ with periodic boundary conditions. The field operator satisfies the Klein-Gordon equation
\begin{equation}
    \Box \hat{\phi}(t,x) = 0,\quad x\in[0,L],
    \label{eq:KGeqn_cavity}
\end{equation}
where the d'Alembertian operator is defined as $\Box=\eta^{\mu\nu}\partial_\mu \partial_\nu$. The field operator can be split into an oscillator-mode contribution $\hat\phi_{\text{osc}}(t,x)$ and a zero-mode contribution $\hat\phi_{\text{zm}}(t,x)$ as
\begin{equation}
    \hat{\phi}(t,x)=\hat\phi_{\text{osc}}(t,x)+\hat\phi_{\text{zm}}(t).
\end{equation}

The oscillator-mode contribution can be expanded in a countable basis of mode functions and their respective complex conjugates, 
\begin{equation}
    \hat\phi_{\text{osc}(t,x)}=\sum_{n\in\mathbb{Z}\setminus\{0\}} \left[\hat{a}_n u_n(t,x) + \hat{a}_n^\dagger u_n^*(t,x)\right],
    \label{eq:fieldamplitudeincavity}
\end{equation} 
where $\hat{a}_n^\dagger, \hat{a}_n$ are the respective creation and annihilation operators of the $n$-th mode of the field. In particular, the mode functions $u_n(t,x)$ are solutions to the Klein-Gordon equation~\eqref{eq:KGeqn_cavity} for the boundary conditions $u_n(t,0)=u_n(t,L)$, given by
\begin{equation}
    u_n(t,x)=\frac{1}{\sqrt{2\omega_nL}}e^{\ii(k_nx-\omega_nt)}
    \label{eq:cavityModeFunction_P},
\end{equation}
where each mode $n$ has corresponding mode frequency $\omega_n=|k_n|$ and wavevector $k_n = \frac{2\pi n}{L}$. The nonzero modes satisfy the orthonormality relations
\begin{align}
    \langle u_n,u_m \rangle_\text{KG}&=\delta_{nm},\nonumber\\
    \langle u_n^*,u_m^* \rangle_\text{KG}&=-\delta_{nm},\nonumber\\ 
    \langle u_n,u_m^* \rangle_\text{KG}&=0,
\end{align}
with respect to the Klein-Gordon inner product \cite{wald1994quantum},
\begin{equation}
    \langle \psi,\varphi \rangle_\text{KG}=\ii\int_0^L \dd x\left(\psi^*\partial_t\varphi -\varphi\partial_t\psi^* \right).
\end{equation}

The zero-mode contribution can be treated as a non-relativistic free particle on the real line of mass $L$,
\begin{equation}
    \hat\phi_{\text{zm}}(t)=\hat{Q}_S+\frac{t}{L}\hat{P}_S,
    \label{eq:zeromode_fieldOp}
\end{equation}
where $\hat Q_S,\hat P_S$ are respective canonical position and momentum Schr\"{o}dinger picture operators that satisfy the canonical commutation relation $[\hat Q_S,\hat P_S]=\ii \mathds{1}$ and that both commute with all creation and annihilation operators of the field oscillator modes $\hat a_n^\dagger,\hat a_n$ \cite{MartinMartinez2014_ParticleDetZeroMode}. Then from \cite{Tjoa2019_ZeroModeSuperluminalSignal}, the commutator of the field amplitude can be written as a sum of the commutator for the zero modes and the oscillator modes as
\begin{equation}
    [\hat\phi(\mathsf{x}), \hat\phi(\mathsf{x}')]=[\hat\phi_{\text{zm}}(t), \hat\phi_{\text{zm}}(t')]+[\hat\phi_{\text{osc}}(\mathsf{x}), \hat\phi_{\text{osc}}(\mathsf{x}')],
    \label{eq:fieldCommutator}
\end{equation}
where
\begin{equation}
    [\hat\phi_{\text{zm}}(t), \hat\phi_{\text{zm}}(t')]=-\frac{\ii \Delta t}{L},\quad \Delta t=t-t',\label{eq:commutator_zm}
\end{equation}
and
\begin{equation}
\begin{aligned}
 & [\hat\phi_{\text{osc}}(\mathsf{x}),\hat\phi_{\text{osc}}(\mathsf{x}')]\\
 &=\lim_{\epsilon\to0^+}\Biggl[
 -\frac{1}{4\pi}\log\left(1-e^{-\frac{2\ii\pi(\Delta u-\ii\epsilon)}{L}}\right)\\
 &\qquad-\frac{1}{4\pi}\log\left(1-e^{-\frac{2\ii\pi(\Delta v-\ii\epsilon)}{L}}\right)\\
 &\qquad+\frac{1}{4\pi}\log\left(1-e^{\frac{2\ii\pi(\Delta u+\ii\epsilon)}{L}}\right)\\
 &\qquad+\frac{1}{4\pi}\log\left(1-e^{\frac{2\ii\pi(\Delta v+\ii\epsilon)}{L}}\right)\Biggr],
\end{aligned}
 \label{eq:osc_field_commutator}
\end{equation}
for double null coordinates in Minkowski space $u=t-x, v=t+x$ and $\Delta u = u-u',\Delta v=v-v'$. The limit is taken in the sense of distributions.

\subsection{UDW Dynamics with a Zero Mode}
\label{sec:UDWModelZeroMode}

\subsubsection{Detector Response}
\label{sssec:DetectorResponse}
The effects of the zero mode on the behaviour of a UDW detector was thoroughly investigated in \cite{MartinMartinez2014_ParticleDetZeroMode}. The setup is as follows. Consider a pointlike two-level UDW detector linearly coupled to the amplitude of a massless scalar quantum field confined to a cavity of length $L$ with periodic boundary conditions. This is described by the interaction Hamiltonian
\begin{equation}
    \hat H_\text{int}=\lambda\chi(\tau)\hat\mu(\tau)\hat\phi(\mathsf{x}(\tau)),\label{eq:H_int_amplitude}
\end{equation}
where $\lambda$ is the coupling constant, $\chi:\mathbb{R}\to[0,1]$ is the switching function controlling whether the interaction is on or off, $\hat\mu(\tau)$ is the monopole operator given by
\begin{equation}
    \hat\mu(\tau)=\hat\sigma^+e^{\ii\Omega\tau}+\hat\sigma^-e^{-\ii\Omega\tau},
\end{equation}
$\hat\sigma^+=\ket{e}\!\bra{g},\hat\sigma^-=\ket{g}\!\bra{e}$ are the respective raising and lowering operators written in terms of the detector's ground state $\ket{g}$ and excited state $\ket{e}$, $\Omega$ is the energy gap of the detector, and $\mathsf{x}(\tau)$ is the worldline of the detector parametrized by its proper time $\tau$. Then assuming that the total system is initially in a state of the form
\begin{equation}
    \hat\rho_0=\hat\rho_{d,0}\otimes\hat\rho_{\text{osc},0}\otimes\ket{\psi}\!\bra{\psi},
\end{equation}
where $\hat\rho_{d,0}$ is an arbitrary detector state, $\hat\rho_{\text{osc},0}$ is a state of the oscillator modes that is diagonal in the Fock basis, and $\ket{\psi}$ is a pure state of the zero mode, it is shown that the zero-mode contributions can be separated from the oscillator-mode contributions to the final state of the detector after the interaction at order $\lambda^2$. In particular, choosing an initial detector state of the form
\begin{equation}
    \hat\rho_{d,0}=a\ket{g}\!\bra{g}+b\ket{g}\!\bra{e}+b^*\ket{e}\!\bra{g}+(1-a)\ket{e}\!\bra{e},
\end{equation}
where $a\in\mathbb{R},b\in\mathbb{C},(a-\frac{1}{2})^2+\vert b\vert ^2\leq\frac{1}{4}$ and choosing the oscillator modes to initially be in the Fock vacuum state $\hat\rho_{\text{osc},0}=\ket{0}\!\bra{0}$, yields the following key observations:
\begin{enumerate}
    \item When the zero-mode state has $\langle\psi\vert \hat Q_S \vert \psi\rangle\neq0$ or $ \langle\psi\vert \hat P_S \vert \psi\rangle\neq0$, its contribution to the final detector state can be of order $\lambda$, whereas the oscillator-vacuum contribution begins only at order $\lambda^2$.
    \item Although the order $\lambda$ zero-mode contribution can be cancelled by choosing $\ket{\psi}$ such that $\langle\psi\vert \hat Q_S \vert \psi\rangle=\langle\psi\vert \hat P_S \vert \psi\rangle=0$, the order $\lambda^2$ zero-mode contribution cannot be cancelled in the same fashion. This is because the $\lambda^2$ zero-mode contribution contains linear combinations of the expectation values of  $\hat Q_S^2, \hat P_S^2, \hat Q_S \hat P_S, \hat P_S \hat Q_S$. These expectation values cannot vanish simultaneously due to the uncertainty principle.
    \item Consider the detector to have a nonzero gap and to be in a setting where no resonant emission or absorption of field quanta is occurring. Additionally, assume that the wavefunction of the zero mode in the position representation is a Gaussian. Then, the zero-mode contribution can become larger than the oscillator-mode contribution for long enough interactions, although both contributions are small in such a regime.
    
    \item Consider a setup where the detector couples to $\frac{\mathrm{d}}{\mathrm{d}\tau} \hat\phi(\mathsf{x}(\tau))$, the proper-time derivative of the field along its trajectory, instead of to the field amplitude $\hat\phi(\mathsf{x}(\tau))$ as in Eq.~\eqref{eq:H_int_amplitude}. Then, the order $\lambda^2$ zero-mode contribution to the populations of the detector's energy levels is proportional to $\langle\psi \vert \hat P_S^2 \vert \psi\rangle$ alone, and so can be made arbitrarily small with a suitable choice of the zero-mode initial state.
\end{enumerate}

Proper-time derivative coupling also removes the additive infrared ambiguity of the massless scalar Wightman function in $(1+1)$-dimensional noncompact settings, as studied in e.g.,~\cite{JuarezAubry2014_DerivativeDetector}. The cavity result above shows why removing this additive ambiguity does not by itself eliminate the state dependence associated with a dynamical zero mode. Related state dependence occurs in expanding spacetimes. \cite{Toussaint2021_MassiveZeroMode} showed that the choice of state of a massive field's spatially homogeneous mode, which becomes effectively massless in an asymptotic region, can substantially affect an inertial detector's response in spatially compact Milne spacetime.


\subsubsection{Entanglement Harvesting}

The zero mode was also shown to affect the results of the entanglement harvesting protocol in~\cite{Tjoa2020_ETZeroMode}, particularly in studies in the Einstein cylinder~\cite{Lin2016_EntanglementEinsteinCylinder}. Following~\cite{Tjoa2020_ETZeroMode}, consider two two-level UDW detectors linearly coupled to the amplitude of a massless scalar quantum field confined to a cavity of length $L$ with periodic boundary conditions. Contrarily to section \ref{sssec:DetectorResponse}, we consider the more general case of smeared UDW detectors, which can be either extended in space or point-like. Assume that both detectors are at rest with respect to the cavity. Then the interaction Hamiltonian can be written in terms of the global Minkowski time $t$ as
\begin{equation}
    \hat H_I(t)=\hat H_\textsc{a}(t)\otimes\mathds{1}_\textsc{b} + \mathds{1}_\textsc{a}\otimes\hat H_\textsc{b}(t),
\end{equation}
\begin{equation}
    \hat H_\nu(t)=\lambda_\nu\chi_\nu(t)\hat\mu_\nu(t)\int_0^L \dd x F_\nu(x-x_\nu)\hat \phi(t,x),
    \label{eq:2UDW_intHam}
\end{equation}
where for detector $\nu\in\{\textsc{a},\textsc{b}\}$ we have that $\lambda_\nu$ is the coupling strength, $\chi_\nu:\mathbb{R}\to[0,1]$ is the switching function controlling whether the interaction is on or off, $\hat\mu_\nu(t)$ is the monopole operator given by
\begin{equation}
    \hat\mu_\nu(t)=\hat\sigma_\nu^+e^{\ii\Omega_\nu(t)}+\hat\sigma_\nu^-e^{-\ii\Omega_\nu(t)},
\end{equation}
$\hat\sigma_\nu^+=\ket{e_\nu}\!\bra{g_\nu},\hat\sigma_\nu^-=\ket{g_\nu}\!\bra{e_\nu}$ are the respective raising and lowering operators written in terms of the ground state $\ket{g_\nu}$ and excited state $\ket{e_\nu}$, $\Omega_\nu$ is the energy gap of the detector, and $F_\nu(x - x_\nu)$ is the spatial smearing function centered at $ x_\nu$.

Assume an initial state of the form
\begin{equation}
    \hat \rho_0=\ket{g_\textsc{a}}\!\bra{g_\textsc{a}}\otimes\ket{g_\textsc{b}}\!\bra{g_\textsc{b}}\otimes\hat\rho_{\text{zm}}\otimes\hat\rho_{\text{osc}},
\end{equation}
where the initial oscillator modes state has vanishing one-point function
\begin{equation}
    \text{Tr}(\hat\rho_{\text{osc}}\hat\phi_{\text{osc}}(\mathsf{x}))=0.
\end{equation}
Then it can be shown that the Wightman function splits into a zero-mode contribution and an oscillator-mode contribution. The split carries over to the pair of detectors' final state at leading order.

For the case of Gaussian switching and delta smearing, the initial state of the zero mode can be tuned to significantly suppress or enhance leading order negativity of the final state depending on the detector parameters chosen. 


Different behaviour is observed however for a similar setup to above, except considering the detectors to couple to the time derivative of the amplitude of the field. In this case, the zero-mode two-point function is $\langle\hat P_S^2\rangle/L^2$. This second moment can be made arbitrarily small, but making it exactly zero would require the zero mode to be in the zero-eigenvalue eigenstate of $\hat P_S$, which is not normalizable. This suppression by a chosen state differs from the operator-level decoupling established for spatial-derivative coupling in Sec.~\ref{sec:results}, where the zero mode drops out regardless of its state. 

\subsubsection{Ambiguity of the Zero-Mode State}
It is now clear that the leading order final state of a UDW detector depends on the state of the zero mode. Choosing a zero-mode state solely to suppress its contribution to detector observables does not, by itself, justify that choice physically. State prescriptions can, however, be motivated by additional physical input. Ford and Pathinayake relate zero-mode quantization to initial conditions in models with a time-dependent mass \cite{Ford1989_ZeroFrequencyInitialConditions}, while Ref.~\cite{AlonsoSerrano2021_ZeroModeQuantization} provides a geometric prescription. The ambiguity therefore concerns predictions made without a specified zero-mode state or a prescription for selecting one. Fixing the oscillator vacuum alone does not determine the detector dynamics.


\subsection{Causality Violations from Neglecting the Zero Mode}

In \cite{Tjoa2019_ZeroModeSuperluminalSignal}, they consider the usual two-level UDW entanglement harvesting scenario with the amplitude-coupling interaction Hamiltonian given by Eq.~\eqref{eq:2UDW_intHam} for the sender--receiver protocol as follows. Assume that the total system is initially in a state of the form
\begin{equation}
    \hat\rho_0=\hat\rho_\textsc{a}\otimes\hat\rho_\textsc{b}\otimes \hat\rho_{\text{osc}}\otimes\hat\rho_{\text{zm}},
\end{equation}
and that the smearing and switching functions are both compactly supported
\begin{align}
    \text{supp}[\chi_\nu(t)]&=[T_\nu^\text{on}, T_\nu^\text{off}],\label{eq:edu_switching_support}\\
    \text{supp}[F_\nu(x-x_\nu)]&=\left[x_\nu-\frac{\sigma}{2}, x_\nu+\frac{\sigma}{2}\right],\label{eq:edu_spatial_support}
\end{align}
where $\sigma$ is the width of the spatial smearing. Let us consider supports that do not overlap,
\begin{align}
    T_\textsc{a}^\text{off} &< T_\textsc{b}^\text{on},\\
    x_\textsc{a}+\frac{\sigma}{2} & < x_\textsc{b}-\frac{\sigma}{2}.
\end{align}

Then it is shown that the second-order signalling contribution to detector B's reduced state, denoted by $\hat\rho_{\textsc{b},\text{signal}}^{(2)}$, can be written as a sum of separate contributions due to the zero mode versus the oscillator modes
\begin{equation*}
    \hat\rho_{\textsc{b},\text{signal}}^{(2)} = \hat\rho_{\textsc{b},\text{signal,osc}}^{(2)} + \hat\rho_{\textsc{b},\text{signal,zm}}^{(2)}.
\end{equation*}
To analyze the effects of the zero mode, a signalling estimator quantifying the strength of signalling from detector A to B is defined by
\begin{equation}
    \varepsilon( x_\textsc{a}, x_\textsc{b})=\biggr\vert \int_0^L \dd t\int_0^L  \dd t' \chi_\textsc{a}(t) \chi_\textsc{b}(t') \mathcal{C}(t, x_\textsc{a},t', x_\textsc{b}) \biggr\vert,
\end{equation}
where the spatially smeared pull-back of the field commutator Eq.~\eqref{eq:fieldCommutator} is defined by
\begin{align}
    &\mathcal{C}(t, x_\textsc{a},t', x_\textsc{b})\nonumber\\
    &=\int_0^L \dd x \int_0^L \dd x'
    F_\textsc{a}( x- x_\textsc{a}) F_\textsc{b}( x'- x_\textsc{b})\nonumber\\
    &\qquad\times\langle [\hat\phi(t, x), \hat\phi(t', x')] \rangle.
\end{align}
In the case of delta switching and pointlike detectors this estimator shows strong causality violations when the zero-mode contribution is removed. At spacelike separation, the omitted zero mode leaves an oscillator commutator proportional to $\Delta t/L$ (see Eq.~\eqref{eq:commutator_zm}). Thus, despite the fact that microcausality is recovered in the large $L$ limit, signalling decays only linearly with $L$. Such causality violations when neglecing the zero mode also appear in higher dimensional spacetimes with toroidal topology.

\section{Results}
\label{sec:results}

In this section, we compare time- and spatial-derivative couplings to a free massless scalar field on the static $(1+1)$-dimensional cylinder. We first define the detector interactions and establish microcausality of both differentiated oscillator fields using the zero-mode commutator in Sec.~\ref{sec:edu_detector_couplings}. We then examine the logarithmic form of the field-amplitude commutator in Sec.~\ref{sec:edu_amplitude_commutator} and calculate its time and spatial derivatives in Secs.~\ref{sec:edu_time_coupling} and~\ref{sec:edu_spatial_coupling}, respectively. Finally, Sec.~\ref{sec:edu_results_discussion} compares the zero-mode contributions to the two interactions and their two-point functions, distinguishing microcausality from independence of the zero-mode state.

\subsection{Detector couplings and microcausality}
\label{sec:edu_detector_couplings}

We work in the rest frame of the static cavity and consider detectors at rest in this frame. For time-derivative coupling, the interaction Hamiltonian is
\begin{equation}
 \hat H_I^{(t)}(t)=\sum_\nu\lambda_\nu\chi_\nu(t)\hat\mu_\nu(t)\int_0^L\!\dd x\,F_\nu(x)\partial_t\hat\phi(t,x),
 \label{eq:edu_time_interaction}
\end{equation}
whereas for spatial-derivative coupling it is
\begin{equation}
 \hat H_I^{(x)}(t)=\sum_\nu\lambda_\nu\chi_\nu(t)\hat\mu_\nu(t)\int_0^L\!\dd x\,F_\nu(x)\partial_x\hat\phi(t,x).
 \label{eq:edu_spatial_interaction}
\end{equation}
Here $\nu\in\{\textsc{a},\textsc{b}\}$ labels the detectors, and $F_\nu$ is a smooth periodic profile localized around detector $\nu$. The switching functions and profiles are chosen so that the two interaction regions are spacelike separated. The pointlike limit is understood only where the required pullbacks are well defined. 

Let $C(\mathsf{x},\mathsf{x}')=[\hat\phi(\mathsf{x}),\hat\phi(\mathsf{x}')]$ denote the full field commutator, and let $C_{\text{osc}}$ denote the commutator of the oscillator field $\hat\phi_{\text{osc}}$. Equation~\eqref{eq:fieldCommutator} implies
\begin{equation}
 C_{\text{osc}}(\mathsf{x},\mathsf{x}')=C(\mathsf{x},\mathsf{x}')
 +\frac{\ii(t-t')}{L}\mathds{1}.
 \label{eq:edu_commutator_decomposition}
\end{equation}
The residual term is independent of position and linear in the time difference. Therefore
\begin{equation}
 \begin{aligned}
  \partial_t\partial_{t'}C_{\text{osc}}&=\partial_t\partial_{t'}C,\\
  \partial_x\partial_{x'}C_{\text{osc}}&=\partial_x\partial_{x'}C.
 \end{aligned}
 \label{eq:edu_derivative_commutators}
\end{equation}
These are distributional identities. Since differentiation cannot enlarge the support of a distribution, both derivative commutators vanish on spacelike-separated open regions. On the cylinder, that condition is
\begin{equation}
 |t-t'|<d_L(x,x'),\qquad
 d_L(x,x')=\min_{n\in\mathbb Z}|x-x'+nL|.
 \label{eq:edu_cylinder_separation}
\end{equation}
Smearing with test functions supported in such regions preserves this vanishing. The periodic null images are causal connections on the cylinder, not exceptions to microcausality.

\subsection{Commutator of Field Amplitudes}
\label{sec:edu_amplitude_commutator}

The commutator of the oscillator modes, Eq.~\eqref{eq:osc_field_commutator}, depends on the null coordinates $u=t-x$ and $v=t+x$. For two events, their differences are
\begin{equation}
 \Delta u=\Delta t-\Delta x,\qquad
 \Delta v=\Delta t+\Delta x,
 \label{eq:edu_null_differences}
\end{equation}
where $\Delta t=t-t'$ and $\Delta x=x-x'$.  Two events are connected by a null ray on the cylinder when  $\Delta u=n L$ or $\Delta v=n L$ for some $n\in\mathbb Z$. These null-coordinate differences are specific to the $(1+1)$-dimensional calculation. The extension to higher dimensions in Sec.~\ref{sec:arbitrary_dimensions} uses the operator decomposition instead.

In Eq.~\eqref{eq:osc_field_commutator}, the two logarithms that depend on $\Delta u$ have the same form as those that depend on $\Delta v$. To write and differentiate this expression more compactly, set $k=2\pi/L$ and define the function
\begin{equation}
 \begin{aligned}
 F_\epsilon(z)&=-\log\bigl(1-e^{-\ii k(z-\ii\epsilon)}\bigr)\\
 &\quad+\log\bigl(1-e^{\ii k(z+\ii\epsilon)}\bigr),
 \qquad\epsilon>0.
 \end{aligned}
 \label{eq:edu_regulated_kernel}
\end{equation}
Here $z\in\mathbb R$ denotes the argument of the function. Substituting $z=\Delta u$ or $z=\Delta v$ gives the corresponding pair of logarithms in Eq.~\eqref{eq:osc_field_commutator}, with the same positive parameter $\epsilon$ already used there. Both $z$ and $\epsilon$ have units of length. The two exponentials contain the common factor $e^{-k\epsilon}$, so expanding the logarithms gives the convergent series
\[
 F_\epsilon(z)=-2\ii\sum_{n=1}^{\infty}
 \frac{e^{-nk\epsilon}}{n}\sin(nkz).
\]
Thus the commutator can be written as
\begin{equation}
 C_{\text{osc}}=\frac{1}{4\pi}\bigl[F_0(\Delta u)+F_0(\Delta v)\bigr],
 \qquad F_0=\lim_{\epsilon\to0^+}F_\epsilon,
 \label{eq:edu_limiting_kernel}
\end{equation}
where the limit is taken in the sense of distributions.

The principal branches explain why this limit is piecewise linear. At finite $\epsilon$, the arguments of both logarithms have positive real part, $1-e^{-k\epsilon}\cos(kz)>0$. They therefore avoid the principal branch cut along the negative real axis, and $F_\epsilon$ is smooth for real $z$. In the limit, for real $z$ away from the points $z=nL$ with $n\in\mathbb Z$, the difference can be written as
\[
 F_0(z)=\log\bigl(-e^{\ii kz}\bigr),
\]
where the logarithm is on its principal branch. As $z$ increases by $L$, the point $-e^{\ii kz}$ makes one counterclockwise turn around the origin. Each time $z$ passes through $nL$, this point crosses the negative real axis, and its principal argument jumps from $+\pi$ to $-\pi$. The corresponding one-sided limits of $F_0$ therefore differ by $-2\pi\ii$.

For $0<z<L$, this branch choice gives $F_0(z)=\ii kz-\ii\pi$. The jumps repeat this linear segment periodically, producing a sawtooth. Extending the same linear formula beyond this interval without resetting the phase would lose those jumps. On each interval between successive multiples of $L$, differentiating the logarithms gives
\[
 \begin{aligned}
 F_0'(z)&=-\ii k\left[
 \frac{e^{-\ii kz}}{1-e^{-\ii kz}}
 +\frac{e^{\ii kz}}{1-e^{\ii kz}}\right]\\
 &=\ii k,
 \qquad F_0''(z)=0.
 \end{aligned}
\]
These identities hold pointwise between the jumps. Each jump contributes $-2\pi\ii\delta(z-nL)$ to the first derivative. Including these contributions gives the distributional identities
\begin{align}
 F_0'(z)&=\ii k-2\pi\ii\sum_{n\in\mathbb Z}\delta(z-nL),\\
 F_0''(z)&=-2\pi\ii\sum_{n\in\mathbb Z}\delta'(z-nL).
 \label{eq:edu_kernel_support}
\end{align}
Evaluating these derivatives at $\Delta u$ and $\Delta v$ places the distributional contributions at $\Delta u=nL$ or $\Delta v=nL$, precisely the periodic null images. Away from those images, we can use $F_0'=\ii k$ and $F_0''=0$ in the calculations below. This includes every spacelike-separated pair of events on the cylinder.

\subsection{Time Derivative Coupling}
\label{sec:edu_time_coupling}
For particle detector models in which both detectors couple to the time derivative of the amplitude of the quantum field, the relevant commutator is
\begin{equation}
    [\partial_t\hat\phi_{\text{osc}}(\mathsf{x}), \partial_{t'}\hat\phi_{\text{osc}}(\mathsf{x}')]=\partial_t\partial_{t'}C_{\text{osc}}(\mathsf{x},\mathsf{x}').
\end{equation}

Using
\begin{equation}
    \partial_t\Delta u=1, \quad \partial_t\Delta v = 1,
\end{equation}
and
\begin{equation}
    \partial_{t'}\Delta u=-1,\quad \partial_{t'}\Delta v=-1,
\end{equation}
we have that
\begin{align}
 &\partial_t\partial_{t'}C_{\text{osc}}(\mathsf{x},\mathsf{x}')\nonumber\\
 &=-\frac{1}{4\pi}\left[F_0''(\Delta u)+F_0''(\Delta v)\right]=0,
 \label{eq:osc_commutator_time_deriv}
\end{align}
away from the distributional support at the periodic null images. 

To verify microcausality explicitly, we must show that the support of the commutator does not intersect the spacelike region. Substituting Eq.~\eqref{eq:edu_kernel_support} gives the global distributional expression
\[
 \begin{aligned}
 &\partial_t\partial_{t'}C_{\text{osc}}(\mathsf{x},\mathsf{x}')\\
 &=\frac{\ii}{2}\sum_{n\in\mathbb Z}
 \bigl[\delta'(\Delta u-nL)+\delta'(\Delta v-nL)\bigr]\mathds{1}.
 \end{aligned}
\]
Here $\delta'$ denotes the derivative of the Dirac delta distribution with respect to its argument. Its support is the same as that of $\delta$, so a contribution can occur only when $\Delta u=nL$ or $\Delta v=nL$. Using $\Delta u=\Delta t-\Delta x$ and $\Delta v=\Delta t+\Delta x$, either condition implies
\[
 |\Delta t|=|\Delta x+nL|
 \qquad\text{for some }n\in\mathbb Z.
\]
These events are connected by a null ray, which may wind around the circle.

By contrast, spacelike separation on the cylinder requires Eq.~\eqref{eq:edu_cylinder_separation}, namely
\[
 |\Delta t|<\min_{n\in\mathbb Z}|\Delta x+nL|.
\]
This strict inequality excludes every point in the support of the commutator. The commutator therefore vanishes as a distribution throughout the spacelike region. In particular, smearing the two field observables with any smooth test functions whose supports are mutually spacelike separated gives a zero commutator. This establishes microcausality of $\partial_t\hat\phi_{\text{osc}}$. The periodic null images introduce no exception at spacelike separation, since they connect causally related events. Hence
\begin{equation}
    [\partial_t\hat\phi_{\text{osc}}(\mathsf{x}), \partial_{t'}\hat\phi_{\text{osc}}(\mathsf{x}')]=0,
\end{equation}
for spacelike-separated points on the cylinder. The time derivative commutator of the oscillator modes is indeed microcausal.

\subsection{Spatial-Derivative Coupling}
\label{sec:edu_spatial_coupling}
For particle detector models in which both detectors couple to the spatial derivative of the amplitude of the quantum field, the relevant commutator is
\begin{equation}
    [\partial_x\hat\phi_{\text{osc}}(\mathsf{x}), \partial_{x'}\hat\phi_{\text{osc}}(\mathsf{x}')]=\partial_x\partial_{x'}C_{\text{osc}}(\mathsf{x},\mathsf{x}').
\end{equation}

Using
\begin{equation}
    \partial_x\Delta u=-1, \quad \partial_x\Delta v = 1,
\end{equation}
and
\begin{equation}
    \partial_{x'}\Delta u=1,\quad \partial_{x'}\Delta v=-1,
\end{equation}
we have that for one spatial derivative
\begin{align}
 \partial_xC_{\text{osc}}(\mathsf{x},\mathsf{x}')
 &=\frac{1}{4\pi}\left[-F_0'(\Delta u)+F_0'(\Delta v)\right]\label{eq:edu_single_spatial}\\
 &=\frac{1}{4\pi}\left(-\frac{2\pi\ii}{L}+\frac{2\pi\ii}{L}\right)=0,\label{eq:edu_single_spatial_zero}
\end{align}
away from the periodic null images. 
A single spatial derivative already removes the spatially constant residual term in Eq.~\eqref{eq:edu_commutator_decomposition}. This does not however remove the null-supported terms from the global commutator $C$.

With derivatives at both points,
\begin{align}
 &\partial_x\partial_{x'}C_{\text{osc}}(\mathsf{x},\mathsf{x}')\nonumber\\
 &=\frac{1}{4\pi}\Bigl[(\partial_x\Delta u)(\partial_{x'}\Delta u)F_0''(\Delta u)\nonumber\\
 &\qquad+(\partial_x\Delta v)(\partial_{x'}\Delta v)F_0''(\Delta v)\Bigr]\label{eq:edu_spatial_chain_rule}\\
 &=-\frac{1}{4\pi}\left[F_0''(\Delta u)+F_0''(\Delta v)\right]=0,
 \label{eq:edu_spatial_zero}
\end{align}
away from the periodic null images. Notice that the resulting distribution is identical to the time-derivative commutator Eq.~\eqref{eq:osc_commutator_time_deriv}. Its support therefore obeys the same condition $|\Delta t|=|\Delta x+nL|$ for some integer $n$, which is incompatible with spacelike separation on the cylinder. This establishes microcausality of $\partial_x\hat\phi_{\text{osc}}$ by the same explicit argument.

Therefore
\begin{equation}
    [\partial_x\hat\phi_{\text{osc}}(\mathsf{x}), \partial_{x'}\hat\phi_{\text{osc}}(\mathsf{x}')]=0,
\end{equation}
for spacelike-separated points.

\subsection{Discussion}
\label{sec:edu_results_discussion}
For spacelike-separated points on the cylinder, we have
\begin{equation}
    [\partial_t\hat\phi_{\text{osc}}(\mathsf{x}), \partial_{t'}\hat\phi_{\text{osc}}(\mathsf{x}')]=0,
\end{equation}
and
\begin{equation}
    [\partial_x\hat\phi_{\text{osc}}(\mathsf{x}), \partial_{x'}\hat\phi_{\text{osc}}(\mathsf{x}')]=0.
\end{equation}
Thus, despite the fact that the oscillator-mode contribution to the field operator alone is noncausal for amplitude coupling, the residual term $\ii(t-t')/L$ is removed by differentiating at both spacetime points. The spatial derivative makes the cancellation especially transparent, since the left- and right-moving contributions cancel already after one spatial derivative. In contrast, for the time derivative coupling one requires the second derivative to observe a vanishing commutator.

As discussed in Sec.~\ref{sec:review}, a major motivation for considering derivative couplings in spacetimes with compact spatial sections is the well-known ambiguity associated with choosing the state of the zero mode. Unlike the oscillator modes, the zero mode does not admit a Fock vacuum and behaves instead as a quantum mechanical free particle. Consequently, the oscillator vacuum alone does not fix the zero-mode state, and predictions that depend on this state require an additional prescription.

From Eq.~\eqref{eq:zeromode_fieldOp}, one obtains
\begin{equation}
    \partial_t \hat\phi_{\text{zm}}(t)=\frac{\hat P_S}{L},
\end{equation}
so that the detector couples directly to the momentum operator of the zero mode. The corresponding Wightman function is readily calculated to be
\begin{equation}
    \langle\partial_t \hat\phi_{\text{zm}}(t)\partial_{t'} \hat\phi_{\text{zm}}(t')\rangle=\frac{\langle\hat P_S^2 \rangle}{L^2}.
    \label{eq:Wightman_timeDeriv_zm}
\end{equation}
Therefore time-derivative coupling retains the zero-mode two-point function and the zero-mode state dependence through $\langle\hat P_S^2\rangle$. Detector responses and entanglement harvesting therefore generally depend on the chosen zero-mode state, consistent with~\cite{MartinMartinez2014_ParticleDetZeroMode,Tjoa2020_ETZeroMode}.

For spatial-derivative coupling, the zero mode is absent already at the level of the interaction. Since the zero mode is spatially homogeneous,
\begin{equation}
    \partial_x \hat\phi_{\text{zm}}(t,x)=0
\end{equation}
identically, irrespective of its quantum state. Consequently,
\begin{equation}
    \left[\partial_x \hat\phi_{\text{zm}}(t,x),\partial_{x'} \hat\phi_{\text{zm}}(t',x')\right]=0,
\end{equation}
and likewise
\begin{equation}
    \langle\partial_x \hat\phi_{\text{zm}}(t,x)\partial_{x'} \hat\phi_{\text{zm}}(t',x')\rangle=0.
\end{equation}

Thus the spatial-derivative interaction contains no zero-mode operator. Holding the initial reduced state of the detectors and oscillator sector fixed, varying the zero-mode state cannot change detector observables. This decoupling is exact for the stated interaction and does not rely on a perturbative truncation. It is the differentiated oscillator field that is microcausal, even though the oscillator amplitude field by itself is not.

The distinction between time- and spatial-derivative couplings is therefore visible in both the interaction Hamiltonian and the two-point function. For time derivatives, the zero-mode commutator vanishes but the zero-mode two-point function generally does not. For spatial derivatives, both vanish because the differentiated zero-mode operator itself is zero. Causal propagation alone would not establish this independence from the zero-mode state.


To reach the conclusions above, it is important to remember that we are working in the rest frame of the cavity. For a detector moving relative to the cavity, the result depends on which spatial derivative the interaction selects. Coupling to the cavity-frame $\partial_x\hat\phi$ still removes the zero mode along any trajectory. On the cylinder, a detector following $(t(\tau),x(\tau))$ has instantaneous rest-frame spatial derivative $\dot x\partial_t+\dot t\partial_x$, where dots denote proper-time derivatives. Thus, by Eq.~\eqref{eq:zeromode_fieldOp}, coupling to $(\dot x\partial_t+\dot t\partial_x)\hat\phi$ would involve coupling to
\begin{equation}
    (\dot x\partial_t+\dot t\partial_x)\hat\phi_{\text{zm}}(t)=\frac{\dot x}{L}\hat P_S,
\end{equation}
introducing back the dependency on the zero-mode state. 
Note that regardless of the derivative chosen for the coupling, microcausality is preserved.

\Needspace{6\baselineskip}
\section{Extension to arbitrary spatial dimension}
\label{sec:arbitrary_dimensions}

We now consider a free real massless scalar field in $(d+1)$ spacetime dimensions, for any integer $d\geq1$. Let the spatial region $\Sigma$ be a rectangular flat torus with periods $L_i>0$ in the coordinates $x^i$, where $i=1,\ldots,d$, and volume $V=\prod_{i=1}^d L_i$. Equivalently, the field obeys periodic boundary conditions in every spatial direction. Time is the Minkowski time of this static geometry, and all spatial derivatives below are taken at fixed cavity time $t$. This setting retains an isolated, normalizable spatially constant mode in every dimension.

We use the same oscillator--zero-mode decomposition as in Sec.~\ref{sec:review}, now with $\bm x=(x^1,\ldots,x^d)$:
\begin{equation}
 \hat\phi(t,\bm x)=\hat\phi_{\text{osc}}(t,\bm x)+\hat\phi_{\text{zm}}(t).
 \label{eq:edu_d_field_decomposition}
\end{equation}
The single integer mode label $n$ is replaced by the vector of integers $\bm n=(n_1,\ldots,n_d)\in\mathbb Z^d$. Equation~\eqref{eq:fieldamplitudeincavity} then becomes
\begin{equation}
 \hat\phi_{\text{osc}}(t,\bm x)=\sum_{\bm n\in\mathbb Z^d\setminus\{\bm0\}}
 \left[\hat a_{\bm n}u_{\bm n}(t,\bm x)
 +\hat a_{\bm n}^{\dagger}u_{\bm n}^*(t,\bm x)\right],
 \label{eq:edu_d_mode_expansion}
\end{equation}
where the periodic mode functions generalize Eq.~\eqref{eq:cavityModeFunction_P} as
\begin{equation}
 \begin{gathered}
 u_{\bm n}(t,\bm x)=\frac{e^{\ii(\bm k_{\bm n}\cdot\bm x-\omega_{\bm n}t)}}{\sqrt{2\omega_{\bm n}V}},\\
 k_{\bm n}^{i}=\frac{2\pi n_i}{L_i},\qquad
 \omega_{\bm n}=|\bm k_{\bm n}|.
 \end{gathered}
 \label{eq:edu_d_mode_functions}
\end{equation}
Only the vector $\bm n=\bm0$, for which every component vanishes, is excluded from the sum. Modes with some vanishing components remain in the oscillator sector. The excluded mode has zero wavevector in every spatial direction and is therefore spatially homogeneous, just as in the $(1+1)$-dimensional case.

The zero mode again has the dynamics of a quantum free particle, with the spatial volume $V$ replacing the length $L$. In the same canonical notation as in Sec.~\ref{sec:review}, its Hamiltonian and commutation relation are
\begin{equation}
 \hat H_{\text{zm}}=\frac{\hat P_S^2}{2V},\qquad
 [\hat Q_S,\hat P_S]=\ii\mathds{1}.
 \label{eq:edu_d_zero_hamiltonian}
\end{equation}
Here $\hat Q_S$ and $\hat P_S$ have the same meaning as in Eq.~\eqref{eq:zeromode_fieldOp}: they are the canonical Schr\"{o}dinger-picture operators of the homogeneous mode. Its Heisenberg-picture field operator is
\begin{equation}
 \hat\phi_{\text{zm}}(t)=\hat Q_S+\frac{t}{V}\hat P_S.
 \label{eq:edu_d_zero_solution}
\end{equation}
For $d=1$, $V=L$, and this is precisely Eq.~\eqref{eq:zeromode_fieldOp}. For equal periods in $d$ dimensions, $V=L^d$, recovering the zero-mode dynamics derived in Appendix C of Ref.~\cite{Tjoa2019_ZeroModeSuperluminalSignal}. No choice of quantum state has entered this construction.

Writing $\partial_i=\partial/\partial x^i$, spatial differentiation removes this operator identically:
\begin{equation}
 \partial_i\hat\phi_{\text{zm}}=0,\qquad
 \partial_i\hat\phi=\partial_i\hat\phi_{\text{osc}},
 \qquad i=1,\ldots,d.
 \label{eq:edu_d_spatial_identity}
\end{equation}
Consequently, the same equality holds for any spatial directional derivative $D_\nu=\sum_{i=1}^d s_\nu^i\partial_i$, where the real coefficients $s_\nu^i$ specify the direction probed by detector $\nu$ in the static frame. The spatial-derivative interaction in Eq.~\eqref{eq:edu_spatial_interaction} generalizes to
\begin{equation}
 \hat H_I(t)=\sum_\nu\lambda_\nu\chi_\nu(t)\hat\mu_\nu(t)
 \int_\Sigma\dd^d\bm x\,F_\nu(\bm x)D_\nu\hat\phi(t,\bm x),
 \label{eq:edu_d_interaction}
\end{equation}
where $F_\nu$ is the spatial profile of detector $\nu$. We take smooth periodic profiles and smooth switching functions for which the interaction evolution is well defined.

Due to the spatial derivative, replacing $\hat\phi$ by $\hat\phi_{\text{osc}}$ in \eqref{eq:edu_d_interaction} leaves the interaction Hamiltonian unchanged, regardless of the zero-mode state or the separations between detectors. This equality of interactions establishes state independence beyond perturbation theory. Denote the joint detector system by $\mathcal D$. Since the interaction acts trivially on the zero-mode Hilbert space, its interaction-picture evolution has the form
\begin{equation}
 \hat U_I(t)=\hat U_{\mathcal D,\text{osc}}(t)\otimes\mathds{1}_{\text{zm}}.
 \label{eq:edu_d_evolution}
\end{equation}
For an initial state $\hat\rho_0$ of the detectors and full field, define $\hat\rho_{\mathcal D,\text{osc}}=\operatorname{Tr}_{\text{zm}}\hat\rho_0$. The evolved detector state is then
\begin{equation}
 \hat\rho_{\mathcal D}(t)=\operatorname{Tr}_{\text{osc}}\!\left[
 \hat U_{\mathcal D,\text{osc}}(t)\hat\rho_{\mathcal D,\text{osc}}
 \hat U_{\mathcal D,\text{osc}}^{\dagger}(t)\right].
 \label{eq:edu_d_reduced_state}
\end{equation}
Thus all detector observables, including harvested entanglement, depend only on the initial reduced state of the detectors and oscillator sector. Holding that reduced state fixed, changing the zero-mode state cannot change the detector dynamics. This statement also allows initial correlations with the zero mode, since it does not assume that $\hat\rho_0$ factorizes.

Microcausality follows independently from the full field commutator. Writing $\mathsf{x}=(t,\bm x)$, Eq.~\eqref{eq:edu_d_zero_solution} gives
\begin{align}
 C_{\text{zm}}(\mathsf{x},\mathsf{x}')
 &=[\hat\phi_{\text{zm}}(\mathsf{x}),\hat\phi_{\text{zm}}(\mathsf{x}')]\nonumber\\
 &=\frac{t'-t}{V}[\hat Q_S,\hat P_S]
 =-\frac{\ii(t-t')}{V}\mathds{1}.
 \label{eq:edu_d_zero_commutator}
\end{align}
The zero mode commutes with the oscillator sector, so $C=C_{\text{osc}}+C_{\text{zm}}$. With $\partial_i=\partial/\partial x^i$ and $\partial_{j'}=\partial/\partial x'^j$, we obtain
\begin{align}
 \partial_i\partial_{j'}C_{\text{osc}}&=\partial_i\partial_{j'}C,
 \qquad i,j=1,\ldots,d,\\
 \partial_t\partial_{t'}C_{\text{osc}}&=\partial_t\partial_{t'}C.
 \label{eq:edu_d_causal_derivatives}
\end{align}
Both identities hold as distributions. Note that the full commutator is supported on causally related pairs of events, and differentiation does not enlarge that support. Both differentiated oscillator commutators therefore must vanish at spacelike separation for these equalities to hold. On this torus, the spacelike condition is
\begin{align}
 |t-t'|&<d_\Sigma(\bm x,\bm x'),\\
 d_\Sigma(\bm x,\bm x')&=
 \min_{\bm m\in\mathbb Z^d}
 \left[\sum_{i=1}^d(x^i-x'^i+m_iL_i)^2\right]^{1/2}.
 \label{eq:edu_d_spacelike_condition}
\end{align}
This is the shortest spatial distance after all periodic identifications. Smearing the derivative observables over spacelike-separated interaction regions preserves the vanishing commutator. No assumption about whether the commutator has support inside the light cone is needed. In particular, the null-image formula of the $(1+1)$-dimensional calculation is not needed for this proof.

The difference between spatial and time derivatives also survives in every dimension. Equation~\eqref{eq:edu_d_zero_solution} implies $\partial_t\hat\phi_{\text{zm}}=\hat P_S/V$, so its contribution to the time-derivative two-point function is
\begin{equation}
 \bigl\langle\partial_t\hat\phi_{\text{zm}}(\mathsf{x})\,
 \partial_{t'}\hat\phi_{\text{zm}}(\mathsf{x}')\bigr\rangle
 =\frac{\langle\hat P_S^2\rangle}{V^2},
 \label{eq:edu_d_time_two_point}
\end{equation}
whenever this expectation value exists. The zero-mode time derivative therefore has a vanishing commutator but a generally nonzero two-point function. Spatial derivatives annihilate the zero-mode operator, preserving microcausality and eliminating zero-mode state dependence in every spatial dimension.

The argument also applies to a static rectangular cavity with homogeneous Neumann boundary conditions on every wall. To see why, let $f_0$ be a zero eigenfunction of the spatial Laplacian. Integration by parts gives
\begin{equation}
 \int_\Sigma\dd^d\bm x\,|\nabla f_0|^2
 =-\int_\Sigma\dd^d\bm x\,f_0^*\nabla^2f_0=0,
 \label{eq:edu_d_neumann_zero}
\end{equation}
where the boundary term vanishes by the Neumann condition. Since the cavity is connected, $f_0$ is constant. Its dynamics and normalization are therefore again given by Eqs.~\eqref{eq:edu_d_zero_hamiltonian} and~\eqref{eq:edu_d_zero_solution}, with $V$ the cavity volume. The state-independence proof is unchanged. The full Neumann field has causal propagation, including reflections at the boundary, so the derivative commutators are again microcausal. For this convex cavity, spacelike separation is $|t-t'|<|\bm x-\bm x'|$, without periodic identification, and the detector profiles can be taken to have support in the interior. The extension established here thus concerns spatially homogeneous zero modes in static finite-volume cavities. It does not require a preferred zero-mode state, but it does require the specified field theory and boundary conditions.

\Needspace{6\baselineskip}
\section{Conclusions}
\label{sec:conclusions}

Zero modes affect detector responses and entanglement harvesting, but their treatment introduces two difficulties. Omitting them from amplitude coupling violates microcausality, while including them requires specifying a state that the oscillator vacuum does not determine. Predictions described as vacuum responses therefore remain ambiguous unless this additional state is physically specified.

We have shown that spatial-derivative coupling resolves both difficulties for detectors stationary in the cavity frame. For a free massless scalar field in static rectangular periodic or Neumann cavities of any spatial dimension, these derivatives annihilate the homogeneous zero-mode operator. The full and oscillator-only interactions consequently coincide. This equivalence holds beyond perturbation theory and allows initial correlations with the zero mode, with the initial reduced detector-oscillator state held fixed. Time-derivative coupling retains a zero-mode operator and therefore does not share this independence from the zero-mode state.

Motion offers a possible way to control this decoupling. On the cylinder, a detector probing its own rest-frame spatial derivative couples to the zero-mode momentum when moving relative to the cavity and decouples when brought to rest. A detector probing the cavity-frame spatial derivative remains decoupled along any trajectory. This distinction suggests investigating controlled motion as a means of preparing, manipulating, and measuring zero-mode states. The ability to switch the zero-mode coupling on and off through detector motion opens new avenues for controlling and probing this degree of freedom. 

In superconducting circuits, the stationary decoupling has a direct physical interpretation. Flux qubits coupled to transmission-line currents cannot probe a spatially homogeneous flux mode, since that mode carries no current~\cite{peropadre2013PRLSuppl,Janzen2023_TunableCouplerTLSWaveguide}. Spatial-derivative coupling thus provides a physically realized interaction for which detector responses and entanglement harvesting require no assumption about the zero-mode state.

\begin{acknowledgments}
EMM acknowledges support through the Discovery Grant Program of the Natural Sciences and Engineering Research Council of Canada (NSERC). IJM acknowledges support from the Natural Sciences and Engineering Research Council of Canada (NSERC) through the Canada Graduate Research Scholarship – Doctoral (CGRS D) as well as from the Government of Ontario through the Ontario Graduate Scholarship (OGS). ATB acknowledges the support of the Natural Sciences and Engineering Research Council of Canada (NSERC), [funding reference number CPRA-610850-2026].
\end{acknowledgments}





\twocolumngrid
\bibliography{references.bib}

\end{document}